# On the Convergence of Collaboration and Knowledge Management

Nesrine Ben yahia, Narjès Bellamine and Henda Ben Ghézala
RIADI-GDL Laboratory
National School of Computer Sciences (ENSI)
University of Manouba, Tunisia
{nesrine.benyahia, narjes.bellamine}@ensi.rnu.tn

*Abstract*—Collaboration technology typically focuses on collaboration and group processes (cooperation, communication, coordination and coproduction). Knowledge Management (KM) technology typically focuses on content (creation, storage, sharing and use of data, information and knowledge). Yet, to achieve their common goals, teams and organizations need both KM and collaboration technology to make that more effective and efficient. This paper is interested in knowledge management and collaboration regarding their convergence and their integration. First, it contributes to a better understanding of the knowledge management and collaboration concepts. Second, it focuses on KM and collaboration convergence by presenting the different interpretation of this convergence. Third, this paper proposes a generic framework of collaborative knowledge management.

*Index Terms*—collaboration, knowledge management, groupware, design.

## I. INTRODUCTION

ORGANIZATIONS increasingly see their intellectual capital as strategic resources that must be managed effectively to achieve competitive advantage. This capital consists of the knowledge held in the minds of its members, embodied in its procedures and processes, and stored in its repositories. Current groupware (cooperative systems) do have the potential to manage knowledge and ensure the knowledge creation, knowledge organization, knowledge sharing, and knowledge use/reuse [1].

Thus, to achieve their goals, people working together must have effective and efficient collaboration processes, and they must be able to bring the intellectual capital of their organization to bear on their task. Subsequently, it should be useful for KM and Collaboration systems to integrate both kinds of capabilities into a single collaborative-and-knowledge system to support joint efforts towards a goal.

This paper examines knowledge management at the collaborative level and is organized as follows; we start with a brief overview of the literature on knowledge management and collaboration. Section three describes different interpretations of knowledge management and collaboration convergence. Finally, sections four presents our generic framework of collaborative knowledge management.

## II. KNOWLEDGE MANAGEMENT

To facilitate the understanding of knowledge management construct, we start our study in this section by defining the term knowledge. Then, we explain the term knowledge management. Finally, we introduce the different levels and approaches of knowledge management.

### A. Knowledge

Knowledge is a somewhat elusive concept [2] having many different definitions. For example, [3] describes knowledge under five different perspectives: state of mind, object, process, access to information, and capability. Knowledge is considered as the sum of information in the context that is dependent on the social group creating it [4]. Knowledge includes information, ideas, and expertise relevant for tasks performed by individuals, groups, or organizations [5]. It is simply introduced as actionable information [6].

According to [7] there are two classes of knowledge: explicit knowledge which '*can be expressed in words and numbers and can be easily communicated and shared in the form of hard data, scientific formulae, codified procedures or universal principles*' and tacit knowledge which is '*highly personal and hard to formalize*'.

### B. Knowledge Management (KM)

Knowledge management is largely regarded as a process involving various activities. Following a literature study of KM practices, [8] synthesize generic KM activities as follow: Create (develop new understandings from patterns and relationships in data, information, and prior knowledge), Collect (acquire and record knowledge), Organize (establish relationships and context so that collected knowledge can be easily accessed), Deliver (search for and share knowledge), Use (bear knowledge on a task).

### C. Knowledge Management levels

[9] summarizes three levels of KM: the individual level (which focuses on the exchange of knowledge between individual workers, whether on the same team or not), the team level (which focuses on the interactions of team members as they collaboratively work together to evaluate information and manage knowledge) and the organizational level (which focuses on the mechanisms that can support and facilitate the distribution of knowledge across an organization of individuals).

*D. KM approaches*

In different viewpoints on KM, many classifications of KM approaches can be distinguished, among them we keep the classification which is originally proposed by [10] and recently adopted by [11] and [12]. These authors distinguish two approaches of KM: codification versus personalization.

Codification approaches consider that Knowledge can be articulated, codified and disseminated in the form of documents, drawings, best practices, etc. Learning can be designed to remedy knowledge deficiencies through structured, managed, scientific processes. [12] considers them as product oriented approaches because they consider knowledge as a product that can be captured and manipulated like any other resource.

Personalization approaches consider that Knowledge is personal in nature and very difficult to extract from people and must be transferred by moving people within or between organizations. Learning must be encouraged by bringing the right people together under the right circumstances. [12] considers them as process oriented approaches because they consider the knowledge creation process between individual as the subject of management.

## III. CSCW AND GROUPWARE

*A. Collaboration*

Collaboration may be seen as the combination of communication, coordination and cooperation [13] and [14]. Communication is related to the exchange of messages and information among people, coordination is related to the management of people their activities and resources, and cooperation is related to the production taking place on a shared space.

*B. CSCW*

Computer Supported Cooperative Work (CSCW) is considered as an attempt to understand the nature and characteristics of cooperative work [15]. It indicates the scientific study and theory of how people work together, how the computer and related technologies affect group behavior, and how technology can best be designed and built to facilitate group work [13].

*C. Groupware*

Groupware is presented as a collection of processes and intentional procedures of group and a collection of software designed to support and facilitate the communication, the coordination and the co-production of group's members [16]. Groupware applications denote any type of software application designed to support groups of people in communication and collaboration on shared information objects [13].

## IV. ON THE CONVERGENCE OF KM AND COLLABORATION

KM and collaboration are complementary [17]. [8] argue that they have common, mutually interdependent purposes and practices and demonstrate the mutual interdependence of knowledge management and collaboration by mapping collaboration technologies to knowledge management activities : *'Technologies for knowledge management may enable improved capture and conveyance of understanding that might otherwise be inaccessible in isolated pockets; technologies collaboration may enable communication and reasoning among people who use knowledge to create value'.*

Following a literature study on KM and collaboration convergence, we identify several terms that are used to denote this convergence: collaborative knowledge management, collaborative knowledge building, collaborative knowledge sharing, collaborative knowledge construction and collaborative knowledge creation.

*A. Collaborative knowledge management (CKM)*

The CKM is considered as a process of collective resolution of problems where it is useful to memorize the process of making collective decision and to structure the group interactions to facilitate problem solving and sharing of ideas [18].

[17] introduces the concept of intellectual bandwidth as the sum of collaborative information system and knowledge management system capabilities within the organization: *"We define Intellectual bandwidth as a representation of all the relevant data, information, knowledge and wisdom available from a given set of stakeholders to address a particular issue".* The proposed Intellectual Bandwidth has three dimensions: the content, the collaboration level and the group size. The content can be data, information, knowledge or wisdom. The collaboration level can be collective (where efforts toward organizational goals are individual and uncoordinated and processes are individualized from start to finish), coordinative (where efforts are coordinated and processes are sequential) or concerted (where efforts and processes are concerted and either simultaneous or asynchronous).

CKM consists of a new community-based collaborative approach to create and share knowledge [19] where two significant aspects have been considered: (a) the internal processes of collaborative knowledge creation and sharing; (b) the effective design of human-computer interfaces facilitating the internal processes, by providing functionalities for the knowledge workers to comprehend, conceptualize, and cooperate in knowledge creation and sharing through e-collaboration processes.

*B. Collaborative knowledge building (CKB)*

The concept of CKB was introduced by [20] where they proposed that schools should function as knowledge building communities in which the construction of knowledge is supported as a collective goal.

[21] presents CKB as a sequential social process in which participants' co-construct knowledge through social interactions and that incorporates multiple distinguishable phases that constitute a cycle of personal and social knowledge building. The cycle of personal understanding starts with building personal comprehension, building tacit pre-understanding and finally constructing personal belief. Then it is possible to enter into an explicitly social process and create new meanings collaboratively. To do this, personal belief is typically articulated in words and expressed in public statements. The cycle of social knowledge-building constructs

argumentation, shared understanding, collaborative knowledge and finally cultural artifacts.

CKB is described also as a group activity where participants use their shared understanding to collaboratively build knowledge in the form of artifacts that are of importance and used in other activities [22].

### C. Collaborative knowledge sharing (CKS)

[23] presents CKS systems as groupware applications that support the development of a shared knowledge repository and shows how conflicts and divergent opinions are an important source to aliment it and their resolution generates new collaborative knowledge. The knowledge repository building is described as a spiral process where knowledge moves from individual knowledge contexts to the community one and comes back to individuals again and is converted from tacit to explicit knowledge.

### D. Collaborative knowledge construction (CKC)

The term CKC is used in [24] to mean a learning process where collaborative groups built on the new ideas offered by others, expressing agreement, disagreement, and modifying the ideas being discussed.

### E. Collaborative knowledge creation (CKC)

Collaborative knowledge creation is defined as the ability to increase the knowledge base or repository, to develop new capabilities and to enhance existing capabilities through combination and knowledge exchange [25]. Combination is the means through which new knowledge is derived through incremental changes to old knowledge. Exchange may involve communication of explicit knowledge in the form of ideas or communication of tacit knowledge in the form of collaborative activities, which can be considered as learning through shared experiences.

## V. TOWARDS A GENERIC FRAMEWORK OF CKM

In this section, we illustrate our reflection about the collaborative knowledge management systems specification by identifying and analyzing the required functionalities of these systems. Then, we expose our generic framework that permits the construction of CKM systems regardless of their properties and their application domain.

### A. CKM specification

Our aim here is to give a functional specification of CKM systems and to identify the cooperative functionalities expected from these systems that facilitate the realization of the collective goal in an effective and efficient manner.

As said about collaborative systems, we start with the required functionalities regarding the collaborative aspect. As it is seen with [13] collaboration must offer three majors functions (communication, coordination and cooperation). [26] argues that to create a shared context and to anticipate actions and requirements related to their collaboration goals, individuals seek the awareness information necessary.

Thus, CKM systems must offer these functions and integrate the knowledge management aspect. On the subject of communication, [18] argues that while communicating, people negotiate and make collective decision, so it is useful to structure the group interactions to facilitate problem solving and sharing of ideas. In fact, the success of communication in the collaboration paradigm entails the understanding of the message by the receiver [26]. Therefore, we point out the following functionality:

*F1: communication and interaction among actors must be structured and "formalized".*

On the subject of coordination, the need for renegotiating and for making collective decisions about non-expected situations that appear during the shared space require a new round of communication, which will require coordination to reorganize the tasks to be executed during cooperation [26]. This coordination may be achieved by social or technical protocol. Therefore, we point out the following functionality:

*F2: cooperative tasks must be planned and coordinated.*

On the subject of cooperation, [26] argues that cooperation is the joint operation during a session within a shared workspace. Group members, within CKM systems, cooperate by producing, manipulating and organizing knowledge, and by building and refining new collective knowledge. Therefore, we point out the following functionality:

*F3: a shared workspace must be offered to facilitate the knowledge management at the collective level.*

On the subject of awareness, participants must be conscious and may obtain feedback from their actions and from the actions of their companions by means of awareness Meta knowledge related to the CKM process. Awareness Meta knowledge consider who (participants), what (collective knowledge), how (management manner), when and where (time and space) of this process. Therefore, we point out the following functionality:

*F4: shared workspace must be enhanced by awareness Meta knowledge.*

Concerning the collaborative knowledge management aspect, [18], [24] and [19] use tacit KM approaches (personalization) to manage the collaborative problem resolution and consider the social process involved in the generation of the collective decision. On the contrary, other authors use explicit KM approaches (codification) to manage the collective knowledge and consider the development of knowledge base in collaboration. Different terms are introduced and used to define this base: [27] uses the term organisational memory, [17] introduce the concept of an intellectual bandwidth, [23] use the concept of a shared knowledge repository.

In our case, as the two types of knowledge (explicit and tacit) can be managed and shared, the two strategies of KM (codification and personalization) must be integrated within the same CKM system: a) the social process and the group tasks must be considered and controlled to enhance the collaborative management of tacit knowledge (such as collective decision), b) the development and the creation of the knowledge base must be considered to enhance the collaborative management of explicit knowledge (such as experience). In addition, we join the idea of [23] on managing two types of knowledge repository: private and shared. In a private context and workspace, individual can administer a private knowledge memory which is only accessible by him and represents the private view of the shared one. In the public context and workspace, there is a unique and shared

knowledge memory which is accessible to everyone. So, we point out the following functionalities:

*F5*: personalization approaches must be considered and enhanced by controlling the social process.

*F6*: codification approaches must be considered and enhanced by managing private and shared knowledge base.

B. *The generic framework of CKM*

After studying the specification of CKM systems, we propose here our generic framework that integrates the knowledge management and collaboration capabilities. The term generic concerns the functional aspect of CKM systems, so the proposed framework represents an attempt to cover the maximum of functionalities required in these systems.

Our framework is inspired from the generic model proposed in [28] to design cooperative systems (groupwares) which is based on three entities: a) cooperation entity that contains the shared software resources, b) coordination entity that contains protocols governing the cooperative tasks and c) actors entity that contains the representation of users within the cooperative model.

By integrating the generic cooperative model of [28] and our CKM specification, we propose a generic framework that permits the management of collaboration, collective knowledge and actors. Thus, it includes three spaces:

*Collaboration Space*: concerns the management of cooperative tasks and it covers the communication, coordination, coproduction and awareness specified respectively in F1, F2, F3 and F4.

*KM Space*: concerns the management of collective knowledge and it covers the strategies of KM specified respectively in F5 and F6. As we mention previously individuals can manage two types of knowledge memory (private and public) and they can work in private or shared context.

Externalization: when individual store knowledge in his private knowledge memory, this knowledge is converted from tacit to explicit in the private context.

Publication: when individuals make public some externalized knowledge and store them in the shared knowledge memory so knowledge are moved from private to shared context, or when groups add knowledge to the shared memory in the public context.

Internalization: when individuals or groups use knowledge from the shared knowledge memory so knowledge is converted from explicit to tacit knowledge however it is moved from shared to private context only for individual internalization.

*Actors Space*: concerns the management and the representation of the different actors and their roles. There are three types of actors:

Individuals: that work independently in a private context.

Groups: dependent individuals that work together in the shared context and engaged to achieve a common goal.

Organizations: groups that work collectively and collaboratively to achieve the organisational goals.

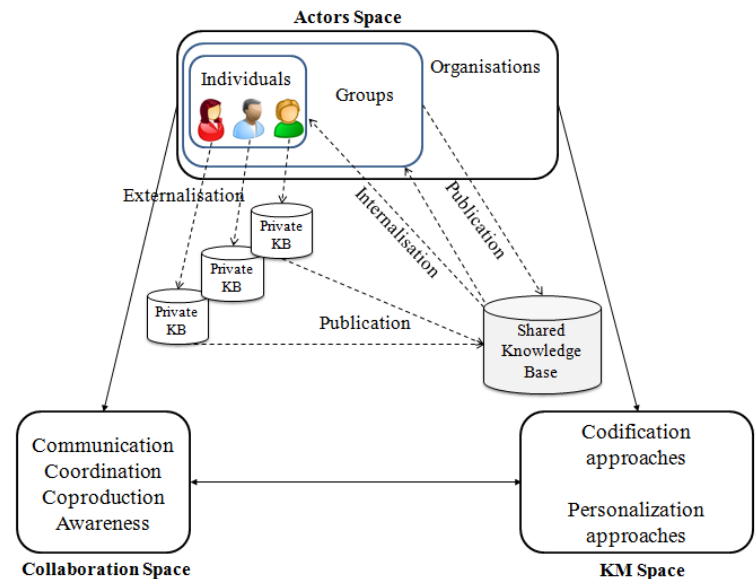

Fig. 1. A generic framework of Collaborative Knowledge management.

## VI. CONCLUSION

By integrating knowledge management and collaboration capabilities, the focus has gradually shifted from collaboration processes and knowledge management processes to Collaborative Knowledge Management processes. This paper focuses on knowledge management and collaboration convergence and analyzes the different functions required of CKM systems. From this specification study, a generic framework of collaborative knowledge management is proposed based on three spaces: collaboration space (to manage and coordinate the cooperative tasks), KM space (to manage the collective knowledge manipulated within the cooperation) and actors' space (to manage and represent the actors implicated into the cooperation).